\documentclass[a4,12pt]{article}
\usepackage{graphics}
\usepackage[usenames]{color}
\usepackage[dvips]{epsfig}
\usepackage{mathrsfs}
\usepackage{amsmath,amssymb}

\newcommand{\norsl}{\normalsize\sl}
\newcommand{\norsc}{\normalsize\sc}
\newcommand{\cs}{\hat{s}}
\newcommand{\ct}{\hat{t}}
\newcommand{\cu}{\hat{u}}
\newcommand{\beq}{\begin{eqnarray}}
\newcommand{\eeq}{\end{eqnarray}}
\newcommand{\la}{\langle}
\newcommand{\ra}{\rangle}
\def\pslash{\rlap/{\mkern-1mu p}}

\begin{document}

\title{\bf 
Soft-fermion-pole contribution to single-spin asymmetry 
for pion production in $pp$ collisions
}

\author{
{\norsc  Yuji Koike and Tetsuya Tomita}
\\
\norsl  Department of Physics, Niigata University,
Ikarashi, Niigata 950-2181, Japan} 
\date{}
\maketitle

%% Abstract 
\begin{abstract}
We study the single transverse spin asymmetry for the inclusive pion production 
in the nucleon-nucleon collision, $p^\uparrow p\to\pi X$,
based on the twist-3 mechanism in the collinear factorization. 
We derive the soft-fermion-pole (SFP) contribution to the twist-3
single-spin-dependent cross section associated 
with the twist-3 quark-gluon correlation functions
in the polarized nucleon.  
We find 
that the SFP can give rise to a large effect to the asymmetry $A_N$
owing to the large partonic hard cross sections with the large color factor,
if the SFP function has a similar magnitude as the soft-gluon-pole (SGP)
function, in spite of the absence of the ``derivative term" for the SFP function
unlike for the SGP function. 
\end{abstract}

\newpage 

The large single transverse spin asymmetry (SSA) observed in $p^\uparrow p\to\pi X$
at FNAL ($\sqrt{S}=20$ GeV)\,\cite{E704}
triggered lots of theoretical and experimental studies
to clarify the origin of the asymmetry. (See \cite{Review} for a review.)
In the conventional framework for the high-energy inclusive reactions, i.e.,
parton model and perturbative QCD, the asymmetry $A_N= (\sigma^\uparrow-\sigma^\downarrow)/ 
(\sigma^\uparrow+\sigma^\downarrow)$ %($\sigma^{\uparrow(\downarrow)}$
%is the cross section with the spin-vector $\uparrow(\downarrow)$ with respect to the
%scatering plane)
is of the order of $\alpha_s m_q/p_T$
where $p_T$ is the transverse momentum of the pion and $m_q$ is the mass of $u$ and $d$
quarks, and hence $A_N$ had been expected to be negligible\,\cite{KPR}.
The transverse polarization of hyperons produced in unpolarized $pp$
collisions was another example of the unexpected large SSAs in early days\,\cite{Lambda}.
More recently the large $A_N$ was also found at higher energy at 
RHIC ($\sqrt{S}=200$, $62.4$ GeV)\,\cite{STAR,PHENIX,BRAHMS}.  HERMES and COMPASS also
reported large SSA
in semi-inclusive deep-inelastic
scattering (SIDIS)\,\cite{hermes,compass}.

Understanding of these SSAs requires the extention of the 
framework for QCD hard processes, and by now the mechanisms leading to such large SSA have been 
understood 
to some extent in terms of these new frameworks, which are often classified into two categories. 
One is the so-called ``T-odd" distribution and fragmentation functions\,\cite{Sivers,Collins}
in the transverse momentum dependent (TMD) factorization approach, and the other is the
twist-3 quark-gluon correlation functions in the collinear factorization 
approach\,\cite{ET82,QS91,EKT07}.
The former approach has had phenomenological successes in the description of SSAs in various 
processes\,\cite{Todd}.  
Universality property of the TMD functions
have been also studied in detail\,\cite{Collins02,BJY03,BMP03,CM04,BMP04}, while 
the factorization proof has been limited to $e^+e^-$,
SIDIS and Drell-Pan processes\,\cite{CS81,CSS85,JMY05}. 
In the framework of collinear factorization SSA is a twist-3 observable, and this 
approach is applicable to wider classes of processes including  
$p^\uparrow p\to\pi X$, although its validity is limited to
large $p_T$ region such that $p_T$ can be regarded as ``hard".  
The formalism of calculating twist-3 single spin dependent cross section with
twist-3 distributions
has been developed in \cite{QS91,EKT07}, in particular, gauge invariance and factorization
property of the cross section was proved in \cite{EKT07} in leading order QCD.
%Although the proof in \cite{EKT07} was shown for SIDIS, it can be applied to other processes
%such as Drell-Yan and $p^\uparrow p\to\pi X$ as well. 
So far the formalism has been applied to SSA in
Drell-Yan\,\cite{QS91,JQVY06,KT071}, 
SIDIS\,\cite{EKT07,EKT06,JQVY06DIS,KT071}, $p^\uparrow p\to\pi X$\,\cite{QS99,KK00,Koike03,Kouvaris,KT072}, 
$pp\to\Lambda^\uparrow X$\,\cite{KK01,hyperon,ZYL08}, heavy-quark production 
in $pp$ collision\,\cite{KQ08,YZ08} etc.
Since SSA is a naively ``T-odd" observable, it appears as an interference between the scattering amplitudes
which have different complex phases.  In the twist-3 mechanism of SSA, this phase
arises as a pole part of a relevant internal propagator in the hard part, and those
poles are classified as soft-gluon-pole (SGP), soft-fermion-pole (SFP) and hard-pole (HP).
It is also shown in \cite{KT071,KT072} that the SGP contribution can be related to a certain
twist-2 cross section in general.  
The connection between the TMD approach and the twist-3 approach at the intermediate region
of the transverse momentum was studied, and it's been shown that
these two mechanisms provide a unique and consistent QCD description of SSA
for the Drell-Yan and SIDIS\,\cite{JQVY06,JQVY06DIS,KVY08}.
Based on these developments, 
we will study in this letter the SSA for $p^\uparrow p\to\pi X$ 
in the framework of the collinear factorization.  

The cross section for $pp\to\pi X$ can be written as a convolution of 
two distribution functions associated with the initial protons, fragmentation function for the final pion,
and the corresponding partonic hard cross sections.  In the twist-3 
cross section,
one of three nonperturbative functions becomes a relevant twist-3 function, and thus
the single-spin-dependent cross section contributing to $p^\uparrow p\to\pi X$
consists of three kinds of twist-3 cross sections: 
\beq
\Delta\sigma^{\rm tw3}&\sim& G^{(3)}(x_1,x_2)\otimes f(x') \otimes D(z)\otimes\hat{\sigma}_A\nonumber\\
&+& \delta q(x)\otimes E^{(3)}(x'_1,x'_2) \otimes D(z)\otimes\hat{\sigma}_B\nonumber\\
&+& \delta q(x)\otimes f(x') \otimes \widehat{E}^{(3)}(z_1,z_2)\otimes\hat{\sigma}_C,
\label{twist3}
\eeq
where the first, second and third factors in each contribution
are, respectively, the polarized distribution function in $p^\uparrow$, unpolarized
distribution in $p$ and the fragmentation function for the pion, and $\hat{\sigma}_{A,B,C}$
are the partonic hard cross sections. 
The two-variable functions $G^{(3)}(x_1,x_2)$, 
$E^{(3)}(x'_1,x'_2)$, 
$\widehat{E}^{(3)}(z_1,z_2)$ are the twist-3 functions, and
$\delta q(x)$ is the transversity distribution in the nucleon.
$f(x')$ and $D(z)$ are, respectively, usual twist-2 unpolarized (quark or gluon) distribution
and fragmentation functions.  
The analysis of the first, second and the third term of
(\ref{twist3}) has been performed in Refs.\cite{QS99,Kouvaris}, \cite{KK00} and \cite{Koike03}, respectively.
In these analyses, only the SGP contributions have been included.  
Kouvaris {\it et al.} derived the complete SGP cross section
for the first term and presented a phenomenological analysis of FNAL and RHIC data\,\cite{Kouvaris}. 
The second contribution was shown to be negligible due to the smallness of 
the hard cross section $\hat{\sigma}_B$\,\cite{KK00}.  This is because only particular diagrams
are allowed to contribute due to the chiral-odd nature of the transversity and $E^{(3)}$,
which remains unchanged by the inclusion of the SFP contribution.
In \cite{Koike03}, the so-called ``derivative" term of the SGP contribution 
from $\widehat{E}^{(3)}(z,z)$
in the third contribution was analyzed, and it was shown that 
this contribution
can be as large as the first one if the nonperturbative function $\widehat{E}^{(3)}(z,z)$
has a similar order of magnitude as $G^{(3)}(x,x)$. 
Recently, however, it's been shown\,\cite{MM08} that the SGP function
$\widehat{E}^{(3)}(z,z)$ becomes identically zero, and thus 
the third term in (\ref{twist3}) requires reanalysis including the
non-partonic pole contribution from (i.e. imaginary part of) $\widehat{E}^{(3)}(z_1,z_2)$.  
In this circumstance, we shall focuss 
in this letter on the analysis of the first contribution in (\ref{twist3}),
in particular, we will derive the complete cross
section formula including the SFP contribution together with the SGP cross section derived in
\cite{Kouvaris}.

There are two independent twist-3 quark-gluon correlation functions (for each quark-flavor $a$),
$G_F^a(x_1,x_2)$ and $\widetilde{G}_F^a(x_1,x_2)$, in the transversely polarized nucleon 
contributing to the first term in (\ref{twist3}).% as $G^{(3)}(x_1,x_2)$. 
They are defined as (see eg. \cite{EKT07})
\beq
M_F^\alpha (x_1,x_2) &=&
    \int \frac{d \lambda}{2\pi} \int \frac{d \mu}{2\pi} e^{i \lambda x_1} 
    e^{i \mu (x_2 -x_1)} \langle PS_\perp |\bar{\psi}_j^a(0) 
    gF^{\alpha \beta}(\mu n) n_{\beta}\psi_i^a(\lambda n) 
    |PS_\perp \rangle  \nonumber\\
         &=& \frac{M_N}{4}(\pslash)_{ij} 
    \epsilon^{\alpha pnS_\perp}G_F^a(x_1,x_2) + 
    i\frac{M_N}{4}(\gamma^5 \pslash )_{ij}S^{\alpha}_\perp 
    \widetilde{G}_F^a(x_1,x_2)+\dots,  
\label{tw3dist}
\eeq
where $M_N$ is the nucleon mass, 
$n$ is a lightlike vector satisfying $p\cdot n=1$, $\epsilon^{\alpha pnS_\perp}\equiv
\epsilon^{\alpha}_{\ \, \mu\nu\lambda}p^\mu n^\nu S_\perp^\lambda$ with $\epsilon_{0123}=1$ and
$+\dots$ denotes twist-4 or higher.  
By $T$- and $P$-invariance, two functions have the symmetry property
under $x_1\leftrightarrow x_2$ as 
\beq
G_F^a(x_1,x_2)=G_F^a(x_2,x_1),\qquad\widetilde{G}_F^a(x_1,x_2)=-\widetilde{G}_F^a(x_2,x_1).  
\eeq
With the gluon's 
field strength $gF^{\alpha \beta}n_\beta$ replaced by the covariant derivative 
$D^\alpha=\partial^\alpha -igA^\alpha$
in (\ref{tw3dist}), one can define another set of twist-3 distributions. The relation
between those functions and the above $\{G_F^a, \widetilde{G}_F^a\}$ 
has been clarified in \cite{EKT06}.  
The twist-3 correlation functions for the ``anti-quark" flavor $G_F^{\bar{a}}$ and 
$\widetilde{G}_F^{\bar{a}}$ can be defined
from (\ref{tw3dist}) by replacing the nonlocal operator
$\bar{\psi}^a(0)gF^{\alpha\beta}(\mu n)n_\beta\psi^a(\lambda n)$
by its charge conjugation\footnote{Here we adopt the convention
that the gluon's field strength $gF^{\alpha \beta}$ is also replaced by its charge 
conjugation (together with $\psi\to\psi^c=C\bar{\psi}^T$) 
when we define the correlation function for the ``anti-quark" flavor.}, 
and is related to the original quark-gluon correlation functions in (\ref{tw3dist}) as
\beq
G_F^{\bar{a}}(x_1,x_2)=G_F^{a}(-x_2,-x_1),\qquad\widetilde{G}_F^{\bar{a}}(x_1,x_2)
=-\widetilde{G}_F^{a}(-x_2,-x_1).  
\label{tw3qbar}
\eeq
For $p^\uparrow p\to\pi X$, SSA occurs from SGP and SFP contributions, which, respectively, set
$x_1=x_2$ and $x_i=0$ ($i=1$ or 2).  Due to the above symmetry property,
$G_F^a$ contributes through SGP and SFP, while $\widetilde{G}_F^a$ contributes only through SFP.
Thus the general structure of the twist-3 cross section takes the following form\footnote{Here
and below we often suppress the flavor
indices from the distribution and fragmentation functions.}:
\beq
  \Delta\sigma^{\rm tw3} &=&    
\biggl(G_F(x,x)-x\frac{dG_F(x,x)}{dx}
    \biggr) \otimes f(x') \otimes D(z) \otimes \hat{\sigma}_{\mathrm{SGP}} \nonumber\\
     &+& G_F(0,x) \otimes f(x') \otimes D(z) \otimes \hat{\sigma}_{\mathrm{SFP}} 
     + \widetilde{G}_F(0,x) \otimes f(x') \otimes D(z) \otimes \hat{\sigma}'_{\mathrm{SFP}},
\label{tw3Xsec}
\eeq
where we have used the fact that
the SGP contribution appears in the combination of $G_F(x,x)-x{dG_F(x,x)\over dx}$\,\cite{Kouvaris,KT072}.
The origin of this combination and the connection of
$\hat{\sigma}_{\mathrm{SGP}}$ to the twist-2 cross section was clarified in \cite{KT072}.
Previous analyses focussed on the SGP term assuming it to be a dominant contribution,
in particular, based on the observation that
the SGP function receives enhancement by the ``derivative" at large $x$\,\cite{QS99}. 
However, there is no clue that the SFP functions $G_F(0,x)$ and $\widetilde{G}_F(0,x)$
themselves are small, and its importance depends
on the behavior of the partonic hard cross sections $\hat{\sigma}_{\mathrm{SFP}}$
and $\hat{\sigma}'_{\mathrm{SFP}}$, even though it does not appear with the ``derivative".
The purpose of this letter is to investigate the SFP contribution 
to $p^\uparrow p\to\pi X$.  We will derive a complete 
twist-3 cross section formula for the SFP contribution and study its impact on
$A_N$ in comparison with the SGP contribution, assuming that
the SFP functions $G_F(0,x)$ and $\widetilde{G}_F(0,x)$
have the same order of magnitude as the SGP function $G_F(x,x)$\,\cite{Koike:2009na}.
\begin{figure}%[H]
  \centering
%\bc
%   \includegraphics[width=14cm,clip]{twist3.eps} 
   \includegraphics[width=14cm]{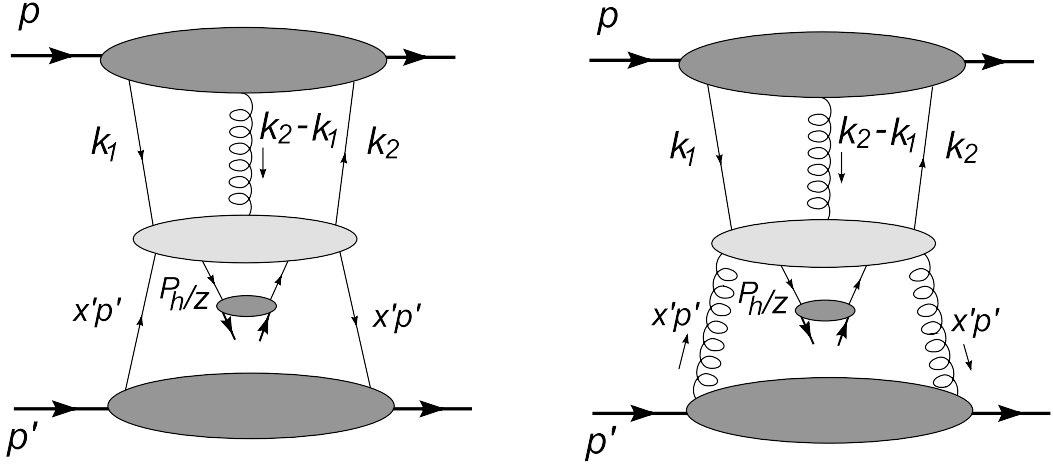} 
%\ec
   \caption{General structure of the cut diagrams for the twist-3 cross section
for $p^\uparrow p\to\pi X$ associated with the twist-3 quark-gluon 
correlation function in the polarized nucleon.
Left (right) diagram represents contribution with the unpolarized quark (gluon) 
distribution in the unpolarized nucleon.} 
   \label{fig1}
\end{figure}

We first recall the formula of calculating twist-3 cross section
for $p^\uparrow (p,S_\perp)+p(p')\to \pi(P_h) +X$ following \cite{EKT07}.  
The contribution to the first term in (\ref{twist3}) arises from the diagrams shown
in Fig. 1, where the distribution function for the unpolarized proton
(bottom blob) and the fragmentation function (middle small blob)
for the pion are already factorized.   
Thus the parton lines coming from the unpolarized proton
have the momentum $x'p'$ and the one fragmenting into the pion have the momentum  $P_h/z$,
which are collinear to the momenta of the parent hadrons.  
The quark and gluon lines with the momenta $k_1$, $k_2$ and $k_2-k_1$ are those 
from the polarized nucleon (top blob).  
The partonic hard parts $S_\rho(k_1,k_2,x'p',P_h/z)$ in Fig. 1 (middle large oval)
are defined for 
the nucleon matrix element $\sim \la pS_\perp| \bar{\psi} A^\rho\psi|pS_\perp \ra$.  
As was demonstrated in \cite{EKT07}, in order to get the gauge invariant twist-3 cross section for SSA,
one needs to reorganize the collinear expansion for $S_\rho(k_1,k_2,x'p',P_h/z)$
with respect to $k_1$ and $k_2$, and one eventually obtains
the twist-3 cross section as
\beq
&&\hspace{-0.8cm}E_h \frac{d^3\Delta\sigma}{d P_h^3} =\frac{\alpha_s^2}{S}
     \int { dx^{\prime} \over x'} \int \frac{dz}{z^2} 
\int dx_1 \int dx_2
   {\rm Tr}\biggl[i\omega^\alpha_{\ \beta}M^\beta_F(x_1,x_2)
   \frac{\partial S_\rho (k_1,k_2,x'p',P_h/z)p^\rho}{\partial
   k^{\alpha}_2}\biggm|_{k_i=x_ip}\biggr]  \nonumber\\
&& \hspace{10cm}\times f(x')D(z),
\label{tw3formula}
\eeq
where 
${\rm Tr}[\cdots]$ denotes trace over spinor and color 
indices\footnote{Color indices connecting
$M_F^\beta$ and $S_\rho$ in (\ref{tw3formula}) are implicit.}, 
$M_F^\beta(x_1,x_2)$ is the correlation function defined in (\ref{tw3dist})
and
$\omega^\alpha_{\ \beta}=g^\alpha_{\ \beta} -p^\alpha n_{\beta}$.  
All the other terms which contributes in the twist-3 level 
%other than that in (\ref{tw3formula}) 
cancel out due to the following two relations\,\cite{EKT07}
\beq
   (x_2-x_1)\frac{\partial S_\rho(k_1,k_2,x^{\prime}p^{\prime},P_h/z)p^\rho}{\partial
    k^{\alpha}_2} \biggm|_{k_i=x_ip}+S_{\alpha}(x_1p,x_2p,
    x^{\prime}p^{\prime},P_h/z) = 0,
\label{ward}
\eeq
and
\beq
  \frac{\partial S_\rho(k_1,k_2,x^{\prime}p^{\prime},P_h/z)p^\rho}
   {\partial k^{\alpha}_1}
   \biggm|_{k_i=x_ip} + \frac{\partial S_\rho (k_1,k_2,x'p',P_h/z)p^\rho}
   {\partial k^{\alpha}_2}\biggm|_{k_i=x_ip} = 0. 
\label{ward2}
\eeq
As noted before, 
single-spin-dependent cross section occurs from the pole (SGP and SFP) contribution
in $\left.{\partial S_\rho (k_1,k_2,x'p',P_h/z)p^\rho\over \partial k_2^\alpha}\right|_{k_i=x_ip}$. 
The pole parts of the sum of the diagrams giving rise to SGP and SFP separately satisfy these relations,
in particular, (\ref{ward}) is a direct consequence of the Ward identity for the pole parts of the diagrams 
(see \cite{EKT07,Tomita09} for the detail).  
For the case of the SFP contribution, which is our main interest in this paper, $x_1\neq x_2$ and 
one can use (\ref{ward}) in (\ref{tw3formula})  to obtain 
\beq
%  \begin{split}
    E_h \frac{d^3\Delta\sigma^{\rm SFP}}{d P_h^3} &=&\frac{\alpha_s^2}{S}
     \int {dx^{\prime}\over x'} \int \frac{dz}{z^2} 
     \int dx_1 \int dx_2\, {1\over x_1-x_2} \nonumber\\
&&\times {\rm Tr}\biggl[i\omega^{\alpha}_{ \ \beta} M^{\beta}_F(x_1,x_2)
S_{\alpha}^{\rm SFP}(x_1p,x_2p,x^{\prime}
   p^{\prime},P_h/z)
\biggr] f(x')D(z),
%  \end{split}
\label{sfpformula}
\eeq
where $S_{\alpha}^{\rm SFP}(x_1p,x_2p,x^{\prime}p^{\prime},P_h/z)$ is the hard part for the SFP 
contribution.  
This formula guarantees absence of the ``derivative" term for the SFP contribution and 
provides a simpler method for the calculation compared with (\ref{tw3formula}).

\begin{figure}%[H]
   \centering
   \includegraphics[width=14cm]{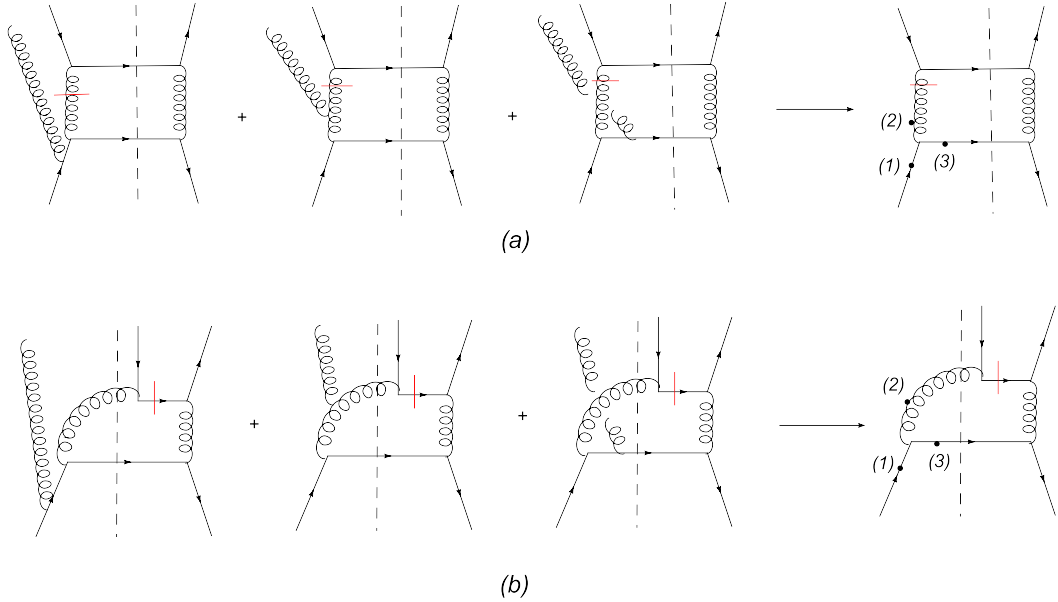} 
   \caption{Typical examples of the cut diagrams for the SFP hard part, $S_\alpha^{\rm SFP}$
in (\ref{sfpformula}).
In (a), two quark-lines coming from the polarized nucleon
are in the opposite side of the cut, while in (b) they are in the same side of the cut.
The bared internal line in each diagram generates SFP.  Momentum for each line
is defined as in Fig. 1 with $k_i=x_ip$. 
In this example, the
coherent gluon line can attach at three different places of each diagram without it,
and those three are represented by a single one shown in the right where 
the positions of the gluon-attachment are represented by the numbered dots (1),(2) and (3).} 
   \label{fig2}
\end{figure} 

To obtain the SFP cross section, one needs to calculate the diagrams with an extra coherent-gluon line
as shown in Fig. 1 with $k_i=x_ip$ ($i=1,\ 2$).  
Those diagrams are classified into two types, typical examples of  which are shown in Fig. 2.
Fig. 2(a) shows the diagrams where the extra gluon line is attached to the cut diagram 
representing a twist-2 hard cross section.  
The line with a bar denotes the propagator generating a SFP. 
For each twist-2 diagram, there are three ways of attaching the coherent gluon,
and thus we denote the sum of those three diagrams by a single diagram
as shown 
in the right side of Fig. 2(a) where 
three dots on the internal lines represent the places to
which the gluon line is attached.  
The mirror diagrams of Fig. 2(a) also give rise to the SFP contribution, and should be included in 
the calculation.  Each of those mirror diagrams give the same result for the SFP cross section
as the original diagrams.

\begin{figure}%[H]
   \centering
   \includegraphics[width=13cm,clip]{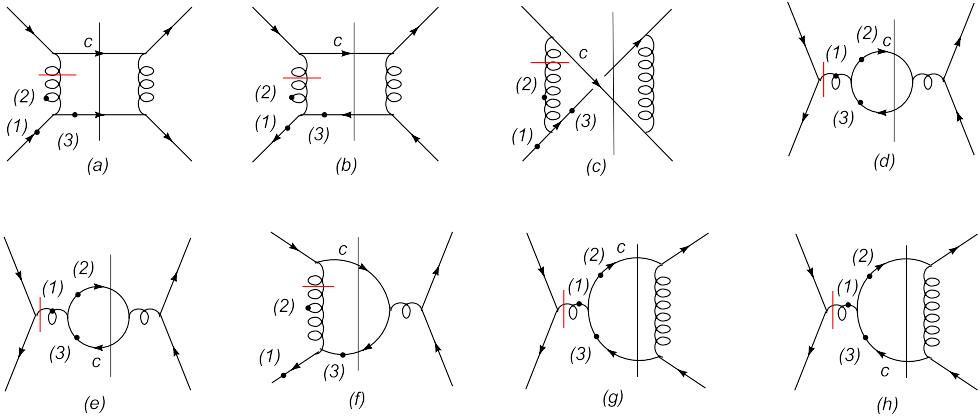} 
   \caption{Diagrams for the SFP contribution of the type shown in Fig. 2(a) in the
quark (or anti-quark) fragmentation channel with the unpolarized quark (or anti-quark)
distribution.  The propagator with a bar gives an SFP.  
The parton $c$ fragments into the final pion.  
The solid vertical line without an arrow represents the final state cut.  The mirror
diagrams also contribute.} 
   \label{fig3}
\end{figure}

The diagrams shown in Fig. 2(b) have the quark and anti-quark lines from the polarized 
proton in the same side of the final-state cut, and the coherent gluon line is attached to the other
side of the cut.  Most of this type of diagrams are obtained by shifting the
position of the final-state cut in the diagrams shown in Fig. 2(a).  For these diagrams, 
there are again three ways of attaching the coherent gluon line, and we denote the sum of those
three by a single diagram with three dots shown in the right side of Fig. 2(b).
For the SFP cross section, some contributions in the same channel from the two types of 
diagrams shown in Figs. 2(a) and (b)
cancel each other, as was the case
for SIDIS\,\cite{KVY08}, which makes the calculation simpler\,\cite{Tomita09}.

\begin{figure}%[H]
   \centering
   \includegraphics[width=11cm,clip]{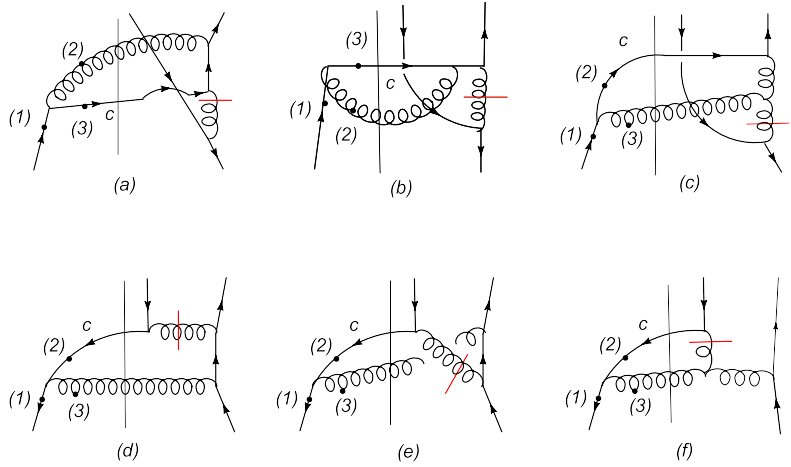} 
 \caption{Diagrams for the SFP contribution of the type shown in Fig. 2(b) in the
quark (or anti-quark) fragmentation channel with the unpolarized quark (or anti-quark)
distribution.  See the caption to Fig. 3.} 
   \label{fig4}
\end{figure} 

Figs. 3$\sim$8 show the diagrams for the SFP contributions
($S_{\alpha}^{\rm SFP}(x_1p,x_2p,x^{\prime}p^{\prime},P_h/z)$ in (\ref{sfpformula})). 
Three or four dots 
numbered as $(1)\sim(4)$ in each diagram represent places where the
coherent gluon is attached as explained in Fig. 2. 
Some diagrams in the figures have 4 dots, since the coupling of the coherent gluon
through the 4-gluon vertex has to be taken into account for those diagrams.
In Figs. 3$\sim$8, we have only shown 
the diagrams which
represent SFP partonic hard cross sections for the ``quark" distribution $G_F^a(0,x)$ and
$\widetilde{G}_F^a(0,x)$ with $x>0$.  By reversing the direction of the arrows
of all the fermion lines in each diagram, one obtain the hard cross sections for
the ``anti-quark" twist-3 distributions defined in (\ref{tw3qbar}). 

\begin{figure}%[H]
   \centering
   \includegraphics[width=13cm,clip]{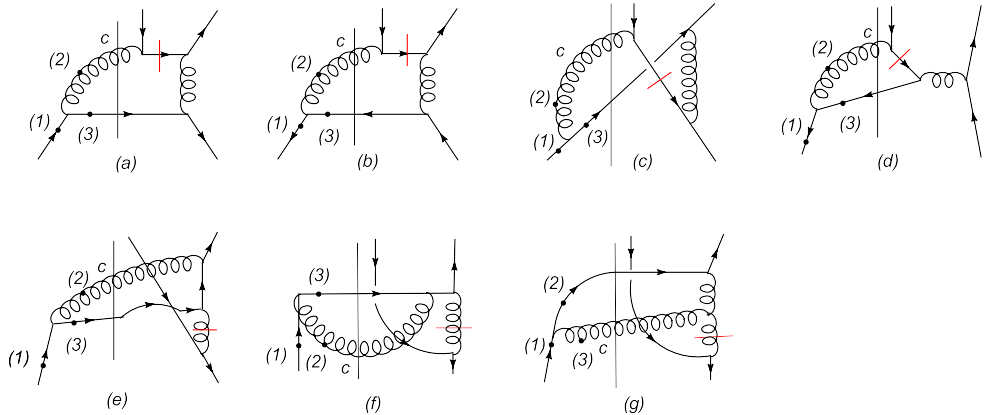} 
 \caption{Diagrams for the SFP contribution of the type shown in Fig. 2(b) in the
gluon fragmentation channel with the unpolarized quark (or anti-quark)
distribution.   See the caption to Fig. 3.} 
   \label{fig5}
\end{figure} 

\begin{figure}%[H]
   \centering
   \includegraphics[width=13cm,clip]{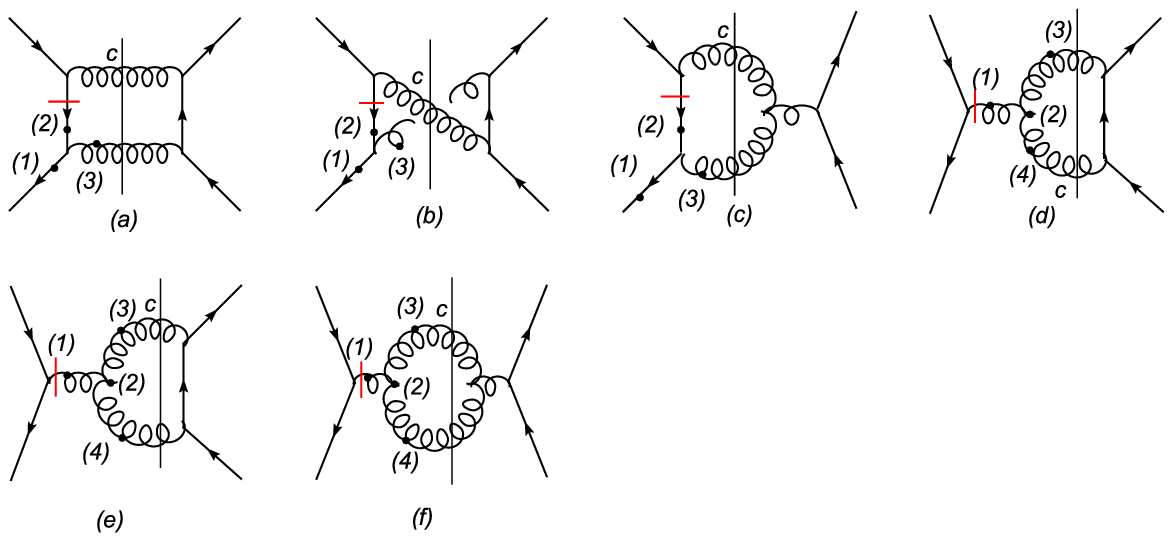} 
 \caption{Diagrams for the SFP contribution of the type shown in Fig. 2(a) in the
gluon fragmentation channel with the unpolarized quark (or anti-quark)
distribution.   See the caption to Fig. 3.} 
   \label{fig6}
\end{figure} 

Figs. 3$\sim$6 represent diagrams in the channels with quark or anti-quark distribution
for the unpolarized nucleon.  Fig. 3 represents diagrams of the type Fig. 2(a) 
and Fig. 4 represents those of the type Fig. 2(b)
in the quark (or anti-quark)
fragmentation channel.
Fig. 5 represents diagrams of the type Fig. 2(b) 
and Fig. 6 represents those of the type Fig. 2(a)
in the gluon 
fragmentation channel.
Fig.~7 represents diagrams 
with the gluon distribution for the unpolarized nucleon and 
quark (or anti-quark) fragmentation function for the pion.  
Fig. 8 represents those
with the gluon distribution for the unpolarized nucleon and 
the gluon fragmentation function for the pion.  
In Figs. 3$\sim$8, we have omitted the diagrams which cancel each other
as was observed for SIDIS\,\cite{KVY08,Tomita09}.

\begin{figure}%[H]
   \centering
   \includegraphics[width=13cm,clip]{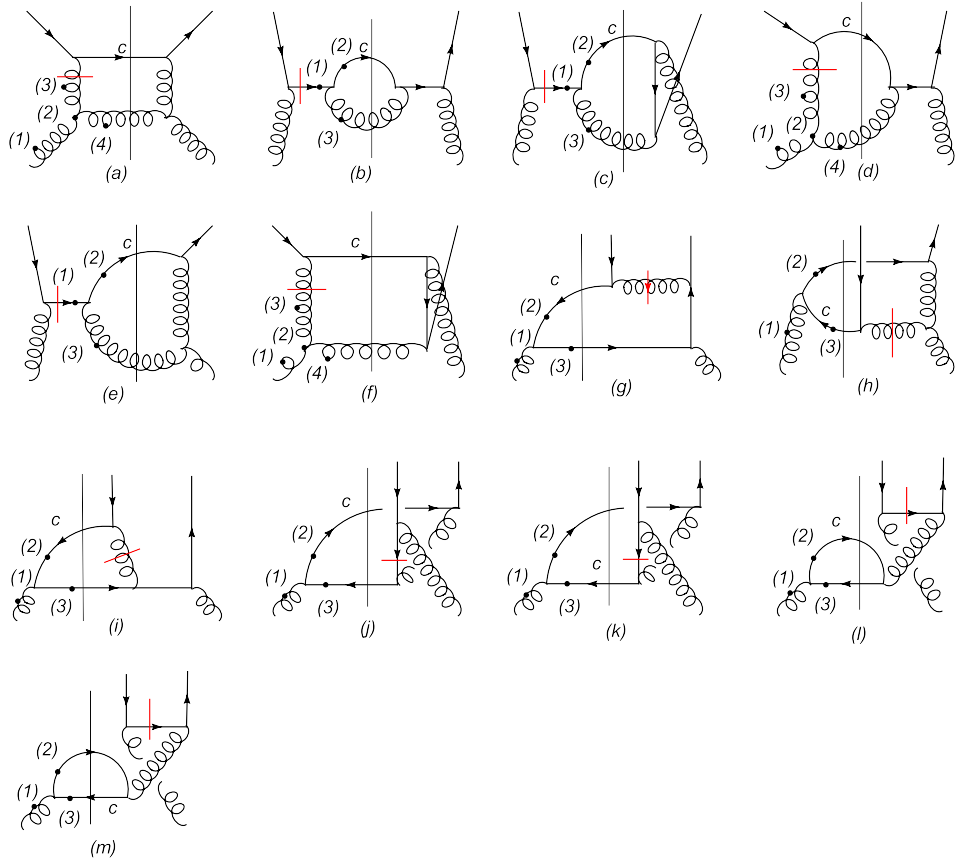} 
 \caption{Diagrams for the SFP contribution in the
quark (or anti-quark) fragmentation channel with the unpolarized gluon 
distribution.   See the caption to Fig. 3.} 
   \label{fig7}
\end{figure} 

\begin{figure}%[H]
   \centering
   \includegraphics[width=12cm,clip]{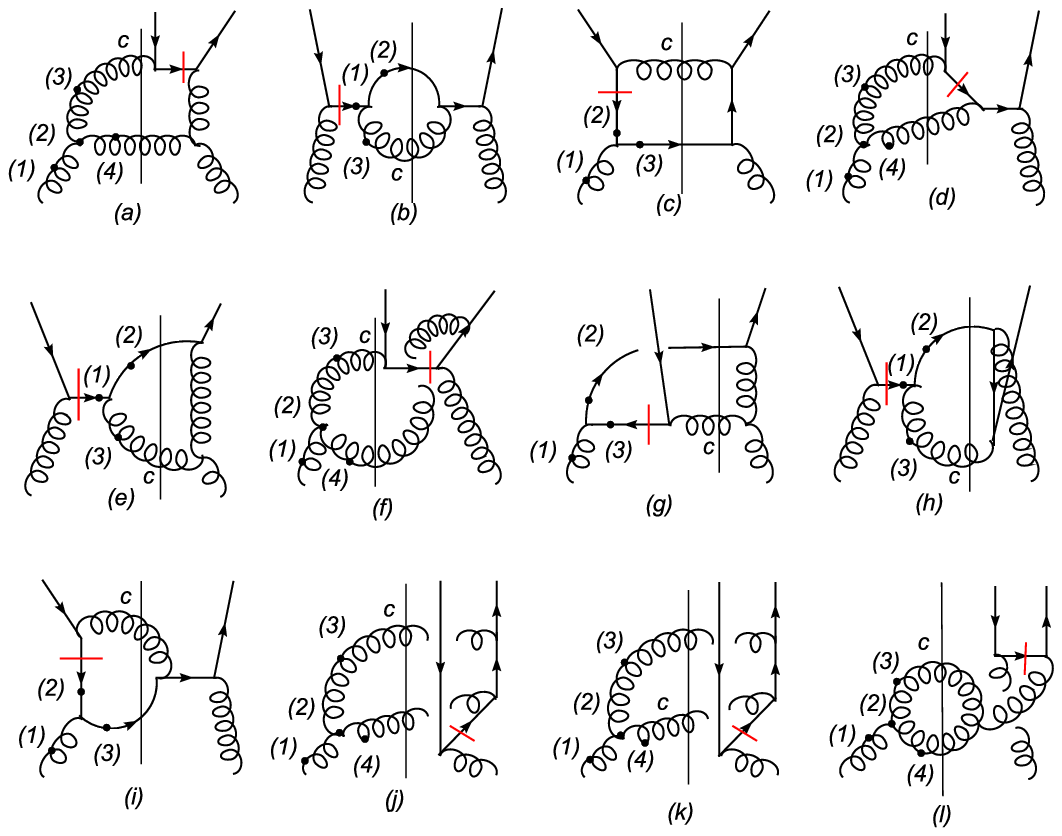} 
 \caption{Diagrams for the SFP contribution in the
gluon fragmentation channel with the unpolarized gluon 
distribution.   See the caption to Fig. 3.} 
   \label{fig8}
\end{figure} 

By calculating all these diagrams, we have eventually obtained the following cross section
for the SFP contribution\,\cite{Koike:2009na,Tomita09}: 
\beq
&&
\hspace{-0.7cm}
E_h \frac{d^3\Delta\sigma^{\rm SFP}}{dP_h^3} =\frac{\alpha_s^2}{S}
   \frac{M_N\pi}{2} \ \epsilon^{pnP_h S_{\perp}} 
\int_{z_{min}}^1
   \frac{dz}{z^3}
 \int_{x^{\prime}_{min}}^1\frac{dx^{\prime}}{x^{\prime}}
 \int  \frac{dx}{x} \,  \frac{1}{x^{\prime}S+T/z}\,\delta \biggl( x-\frac{-x^{\prime}U/z}
     {x^{\prime}S+T/z} \biggr)  \nonumber \\[5pt]
&&\qquad\quad \times \left[
\sum_{a,b,c} \left( G_F^a(0,x) + \widetilde{G}_F^a(0,x) \right) 
\left\{ 
q^b(x')\left( D^c(z) \hat{\sigma}_{ab\to c} +   D^{\bar{c}}(z)\hat{\sigma}_{ab\to \bar{c}} 
\right) \right.\right. \nonumber\\
&&\left.\left. \hspace{6.5cm} + q^{\bar{b}}(x')
\left( D^c(z)\hat{\sigma}_{a\bar{b}\to c} +   
D^{\bar{c}}(z)\hat{\sigma}_{a\bar{b}\to \bar{c}} \right) 
\right\} 
\right. 
\nonumber \\[5pt]
&&\left. 
\qquad\qquad  + \sum_{a,b} \left( G_F^a(0,x) + \widetilde{G}_F^a(0,x) \right) 
\left(
q^b(x')D^g(z)\hat{\sigma}_{ab\to g} + q^{\bar{b}}(x')D^{g}(z)\hat{\sigma}_{a\bar{b}\to g} 
\right) 
\right. \nonumber\\
&&\left. 
\qquad \qquad + \sum_{a,c} \left( G_F^a(0,x) + \widetilde{G}_F^a(0,x) \right) G(x')\left(
D^c(z)\hat{\sigma}_{ag\to c} + D^{\bar{c}}(z)\hat{\sigma}_{ag\to \bar{c}} 
\right) \right.\nonumber\\
&& 
\qquad\qquad \left. + \sum_a\left( G_F^a(0,x) + \widetilde{G}_F^a(0,x) 
\right)G(x')D^g(z)\hat{\sigma}_{ag\to g}
\right],
\label{SFPfinal}
\eeq
where the Mandelstam variables for the process are defined as
$S =(p+p^{\prime})^2 = 2p\cdot p^{\prime}$, $T =(p-P_h )^2 = -2p\cdot P_h$ and
$U =(p^{\prime}-P_h )^2 = -2p^{\prime}\cdot P_h$,
$q^b(x')$ denotes unpolarized quark distribution for the quark flavor $b$,
$G(x')$ denotes unpolarized gluon distribution, and $D^i(z)$ ($i=c$ for the quark-flavor and 
$i=g$ for gluon) is the pion fragmentation function.  
In the summations $\sum_{a,b,c}$, $\sum_{a,b}$, $\sum_{a,c}$
and $\sum_{a}$, the sum of $a$ should be taken over all quark and
anti-quark flavors ($a=u,d,s,\bar{u},\bar{d},\bar{s},\cdots$) and 
the sum of $b$ and $c$ is restricted onto quark flavors when $a$ is a quark and
onto anti-quark flavors when $a$ is an anti-quark.  (For an anti-quark $b$, $\bar{b}$ denotes
quark flavor.)
%x =\frac{-x^{\prime}U/z}{x^{\prime}S+T/z},
The integration region in the convolution formula is specified by
$x^{\prime}_{min} =\frac{-T/z}{S+U/z}$ and $z_{min}= -\frac{(T+U)}{S}$.
$\hat{\sigma}_{ab\to c}$ {\it etc} represents partonic hard cross sections which
are the functions of 
Mandelstam variables in the parton level
$\hat{s}=(xp+x^{\prime}p^{\prime})^2=xx^{\prime}S$, 
$\hat{t}= (xp-P_h/z)^2=\frac{x}{z}T$ and
$\hat{u}=(x^{\prime}p^{\prime}-P_h/z)^2=\frac{x^{\prime}}{z}U$
and are given as follows:
\beq
\hat{\sigma}_{ab\to c} 
&=&{-(N^2\cs+2\ct)(\cs^2+\cu^2)\over N^2 \ct^3 \cu} \delta_{ac}
+ { -(N^2\ct+\cu-\cs)\cs \over N^3\ct\cu^2}\delta_{ab}\delta_{ac},\nonumber\\
\hat{\sigma}_{ab\to \bar{c}} &=&0,\nonumber\\
\hat{\sigma}_{a\bar{b}\to c} 
&=&{(N^2\cu+2\ct)(\cs^2+\cu^2)\over N^2 \ct^3 \cu} \delta_{ac}
+{(N^2\cu+2\cs)(\ct^2+\cu^2)\over N^2 \cs^2\ct \cu} \delta_{ab}
-{ (N^2-1)\cu^2\over N^3 \cs\ct^2} \delta_{ab}\delta_{ac},\nonumber\\
\hat{\sigma}_{a\bar{b}\to \bar{c}} %&=& \hat{\sigma}^7_{(e)} \delta_{ab}
&=&{-(N^2\ct+2\cs)(\ct^2+\cu^2)\over N^2 \cs^2\ct \cu} \delta_{ab}
+ { -N^2\cs+\ct-\cu \over N^3\cu^2} \delta_{ab}\delta_{ac},
\eeq
\beq
\hat{\sigma}_{ab\to g} &=& {(N^2\cs+2\ct)(\cs^2+\cu^2)\over N^2 \ct^3 \cu} 
+{-1\over N^3\cs\ct\cu^2}
\left(N^2(\cs^3+3\cs^2\cu-2\cu^3)+\cs^3-\cs^2\cu\right) \delta_{ab},\nonumber\\
\hat{\sigma}_{a\bar{b}\to g} &=& {-(N^2\cu+2\ct)(\cs^2+\cu^2)\over N^2 \ct^3 \cu} \nonumber\\
&+&\left\{{1\over N^3}\left({\cu\over \cs\ct}+{1\over \cu}\right)
+{1\over N}\left( {\cs^2+\cs\ct+\ct^2\over \cs\cu^2}-{\cu\over \ct^2}\right) + 
{ N(\cu^3-\ct^3)(\ct^2+\cu^2)\over \cs^2\ct^2\cu^2}\right\}\delta_{ab},
\eeq
\beq
\hat{\sigma}_{ag\to c} &=& 
\left\{
{ N^2(\cs^3-\cu^3)(\cs^2+\cu^2) \over (N^2-1)\cs\ct^3\cu^2}
+{ \cs\cu(\cs^2+\cs\cu-\cu^2) - N^2(\cs^4+\cs^3\cu+\cs^2\cu^2+\cs\cu^3+\cu^4)
\over N^2(N^2-1)\cs\ct^2\cu^2} \right\}
\delta_{ac}\nonumber\\
&& +{ (N^2\cu+2\cs)(\ct^2+\cu^2) \over N(N^2-1)\cs^2\ct\cu},\nonumber\\
\hat{\sigma}_{ag\to \bar{c}} &=&
 {\cs+2\ct-N^2\cs \over N^2(N^2-1)\cu^2} \delta_{ac} +
{ -(N^2\ct+2\cs)(\ct^2+\cu^2) \over N(N^2-1)\cs^2\ct\cu},
\eeq
\beq
\hat{\sigma}_{ag\to g} &=&
{ -N^2\over (N^2-1)\cs^2\ct^3\cu^2}
\left( 4\cs^6+11\cs^5\ct +19\cs^4\ct^2+22\cs^3\ct^3+19\cs^2\ct^4+11\cs\ct^5+4\ct^6
\right) \nonumber\\
&&+ { 1\over N^2(N^2-1)\cs\ct^2\cu^2}
\left\{ -\cs\ct\cu^2 +N^2(\cs^4+\cs^3\ct+2\cs^2\ct^2+\cs\ct^3+\ct^4)\right\},
\eeq
where $N=3$ is the number of colors.  
A remarkable feature of (\ref{SFPfinal}) is that the partonic hard cross sections 
for $G_F(0,x)$ and $\widetilde{G}_F(0,x)$ are the same, even though
each diagram in Figs. 3$\sim$8 gives different contributions 
for the two functions.  Accordingly, they appear in
the combination of $G_F(0,x) +\widetilde{G}_F(0,x)$ in (\ref{SFPfinal}).
Although we expect there is a reason for this simple result, we do not understand the origin at this point. 
Another important feature is
that some terms in the hard cross section (the first terms in 
$\hat{\sigma}_{ab\to c}$,
$\hat{\sigma}_{a\bar{b}\to c}$, $\hat{\sigma}_{ag\to c}$ {\it etc.}) 
have the same $1/\ct^3$ behavior as the SGP hard cross section
which gives dominant contribution to $A_N$ in the forward region ($x_F\to 1$).
In addition their color factors are typically $N^2=9$ times larger than those
in the SGP cross section, making the magnitude of the SFP hard cross section huge.  
This suggests that the SFP contribution
gives rise to a significant contribution to the asymmetry at large $x_F$
even though the SFP function does not appear with the derivative.  

To see the impact of the SFP contribution, we have performed 
a numerical calculation of $A_N$ for $p^\uparrow p\to\pi X$, assuming
that the SFP function is of the same order of magnitude as the SGP function.
Kouvaris {\it et al.} \cite{Kouvaris} have parametrized the
SGP function $G_F^a(x,x)$ so that the SGP contribution reproduces
the $A_N$ data obtained from RHIC and FNAL data.  Their analysis shows 
that both data are reasonably well reproduced. In particular, they found that 
the derivative term in (\ref{tw3Xsec})
brings dominant contribution compared to the nonderivative contribution. 
Here we adopt Fit(I) of \cite{Kouvaris} and assume $G_F^a(0,x)+\widetilde{G}_F^a(0,x)=G_F^a(x,x)$
($a=u,d$).  
Fig. 9 shows $A_N$ for the pion at $\sqrt{S}=200$ GeV and $P_{hT}=1.5$ GeV
with and without SFP contribution.  As is seen from the figure that
the SFP contribution brings large effect in the positive $x_F$ region,
while its effect is negligible in the negative $x_F$ region.
This is due to the large color factors for the $1/\ct^3$ terms in
$\hat{\sigma}_{ab\to c}$,
$\hat{\sigma}_{a\bar{b}\to c}$, $\hat{\sigma}_{ag\to c}$ as noted above.  
From this figure it is clear that
the SFP contribution can affect $A_N$ significantly 
even though it does not receive enhancement by the derivative unlike SGP contribution,
unless the SFP function itself is small.  The detail of the present calculation together with
a phenomenological analysis of the data will be presented  elsewhere\,\cite{Tomita09}. 

To summarize, we have calculated the SFP contribution to
the cross section for $p^\uparrow p\to h X$ 
associated with the twist-3 quark-gluon correlation functions in the polarized nucleon.  
We have also shown that its effect is significant and should be included in the analysis
of $A_N$.  
For a more complete analysis, one needs to include the contribution from
the triple-gluon twist-3 distribution function as well.  Recent study\,\cite{KQ08}
shows that $A_N$ for the open charm production 
has a potential to determine the function.  We hope that
the global analysis of all those data including all the QCD effects 
will clarify the origin of observed SSA.

\begin{figure}
\centering
\includegraphics[height=.35\textheight]{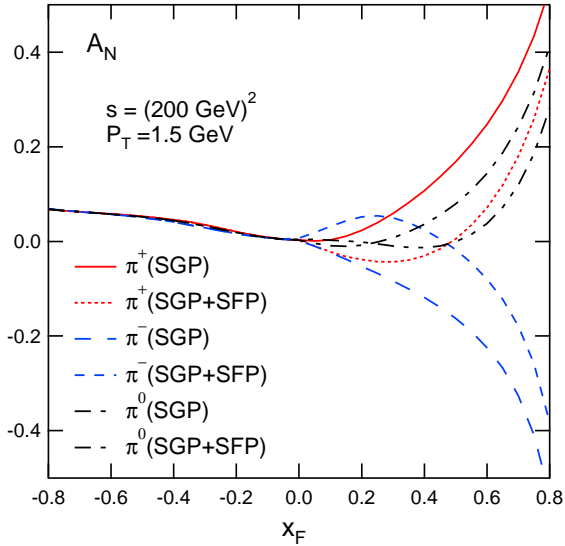}
\caption{$A_N$ for $p^\uparrow p\to\pi X$ at $\sqrt{S}=200$ GeV and $P_{hT}=1.5$ GeV.
Solid, long-dashed, and dash-dot lines are, respectively, $A_N$ for $\pi^+$, $\pi^-$ and $\pi^0$
obtained with only the SGP contribution.  
Dotted, short-dashed, and dash-double-dot lines are, respectively, 
$A_N$ for $\pi^+$, $\pi^-$ and $\pi^0$
obtained with both SGP and SFP contributions.  }
\label{fig9}
\end{figure}

\vspace{0.5cm}

\noindent
{\bf Acknowledgement}

The work of Y.~K. is supported in part by the Uchida Energy Science Promotion
Foundation.

\end{document}